# Automated Creation and Human-assisted Curation of Computable Scientific Models from Code and Text

**Varish Mulwad, Andrew Crapo\*, Vijay S. Kumar, James Jobin, Alfredo Gabaldon, Nurali Virani, Sharad Dixit, Narendra Joshi\***

**GE Research**

{varish.mulwad, crapo, v.kumar1, jobin, alfredo.gabaldon, nurali.virani, sharad.dixit, narendra_joshi}@ge.com

## Abstract

Scientific models hold the key to better understanding and predicting the behavior of complex systems. The most comprehensive manifestation of a scientific model, including crucial assumptions and parameters that underpin its usability, is usually embedded in associated source code and documentation, which may employ a variety of (potentially outdated) programming practices and languages. Domain experts cannot gain a complete understanding of the implementation of a scientific model if they are not familiar with the code. Furthermore, rapid research and development iterations make it challenging to keep up with constantly evolving scientific model codebases. To address these challenges, we develop a system for the automated creation and human-assisted curation of a knowledge graph of computable scientific models that analyzes a model's code in the context of any associated inline comments and external documentation. Our system uses knowledge-driven as well as data-driven approaches to identify and extract relevant concepts from code and equations from textual documents to semantically annotate models using domain terminology. These models are converted into executable Python functions and then can further be composed into complex workflows to answer different forms of domain-driven questions. We present experimental results obtained using a dataset of code and associated text derived from NASA's Hypersonic Aerodynamics website.

## Introduction

Scientific models are often developed and refined to analyze and better approximate various real-world phenomena. In particular, they encapsulate the behavior of complex systems that are governed by physical laws. To truly comprehend and further build upon an existing set of scientific models requires extensive awareness of each model's usability, including the circumstances under which it can and cannot be applied, any underlying assumptions (e.g., with respect to input data or operating conditions), parameters, and so on. While research publications and technical reports remain the prevailing medium for formally disseminating such scientific knowledge, shortcomings of this medium coupled with the increasingly rapid innovation cycles across scientific domains are leading to knowledge gaps whereby one can no longer rely solely on text from scientific literature to be able to discover and apply these models.

A more advanced, comprehensive understanding of a scientific model can potentially be gleaned from its implementation in source code, along with any associated code comments and formal and informal documentation (Rollins et al. 2014). The trend of publishing model implementations is rapidly growing with different domains maintaining their own respective repositories of scientific models. For example, the Community Surface Dynamics Modeling System (CSDMS) (Peckham, Hutton, and Norris 2013) has a collection of nearly 400 models; the CoMSES Net Computational Model Library (Rollins et al. 2014) has over 800 models of social and biological systems; NetLogo Modelling Commons[2] has over 1000 agent-based models, and there are several more models on GitHub – e.g., the Institute for Health Metrics and Evaluation is assembling hundreds of models to study various aspects of the COVID-19 pandemic[3].

Domain experts and scientists face several challenges in interacting with source code artifacts for scientific models. Firstly, a general lack of awareness of software practices or unfamiliarity with specific programming languages could prevent them from tapping into important knowledge otherwise embedded in the code. Secondly, the models could be implemented using legacy and/or obscure programming technologies making them hard to interpret and execute. Models in both CSDMS and CoMSES are programmed in at least eight different languages ranging from legacy (Fortran77, Fortran90, C) to modern (Java, Python). Lastly, they must deal with poorly documented code, and bad code design that makes it cumbersome to follow the execution flow and to see the scientific model in action. While this burgeoning amount of scientific knowledge residing in not so well-maintained codebases poses a serious challenge, little or no attention has been paid by the community to extracting scientific models from software code and associated text sources. Existing efforts have largely focused on extraction of concepts, tasks, methods, and relationships from text alone (Gábor et al. 2018).

---

[*] Work performed by author(s) while at GE Research
[2] http://modelingcommons.org
[3] https://github.com/orgs/ihmeuw/repositories

We address these challenges and shortcomings by developing a novel system for extracting scientific models from code and text sources, aligning them with global ontologies, and automatically populating them into a knowledge graph (KG) of computable models (CM) that are grounded in Semantic Web standards. Our overarching goal is to enable domain experts to easily interact with, curate, and in the future, modify (e.g., customize) the assembled KG of models at the semantic layer. To this end, our contributions include:

a) a rules-based and semantic reasoning-driven approach to parse and extract relevant code artifacts (methods and their inputs, outputs, etc.) associated with a scientific model from code sources,
b) a knowledge- and data-driven approach to further augment the above artifacts with contextual knowledge – such as scientific concepts, equations, and the semantic meaning of the constituent variables (aka 'augmented types') – for the models as extracted from associated text sources, and
c) a user-friendly interface (+ English-like language) for domain experts to correct and curate extracted knowledge.

While these curated scientific knowledge graphs can facilitate automated composition of the executable CMs (towards supporting a variety of downstream model-oriented tasks such as prognosis, prediction, and optimization), the usability of the KGs themselves is not a focus of this paper.

## Approach

Figure 1 presents our system for extracting scientific knowledge from code and text sources. As shown, there are two core modules – the *Code2Triples* module identifies and extracts source code elements and relationships between them, while the *Text2Triples* module extracts scientific concepts, equations, and augmented types from text. Interaction between these modules leads to a collective knowledge extraction for a given model – e.g., comments associated with code elements as identified by *Code2Triples* are processed by *Text2Triples* to identify their semantic meaning in domain terms. Using this knowledge, further elements of interest are extracted by *Code2Triples*. The *Code2Triples* module applies rule-based reasoning over extracted code elements to determine implicit inputs and outputs, constants, dead code, unreachable code, etc. On the other hand, the Text2Triples module further aligns the extracted concepts with known concepts from Wikidata.

[4] https://javaparser.org/

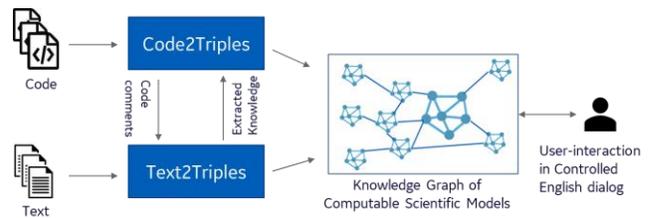

*Figure 1. Knowledge extraction is performed over user provided code and text files. Code is processed by Code2Triples and text files and code comments are processed by Text2Triples.*

A scientific meta-model or ontology is used to represent extracted knowledge from code and text in a knowledge graph. The extracted code elements are translated into their equivalent representation in the Python language to form computable models – these are then made available to users for further curation (via controlled English). Here, *Text2Triples* converts both extracted and user-provided equations into executable Python code. In instances where relationships between, say, variables are not explicitly captured in code, this ability to auto-generate code from equations can help fill important gaps in the knowledge graph. We further describe each module in the following subsections.

### Code2Triples

Extraction of scientific knowledge from code is accomplished by using the standard Open Source JavaParser.[4] The Abstract Syntax Tree (AST) generated by the parser contains all the syntactic elements of the code along with their type and the code structure into which they fit. Comments are also identified and associated in as much as is possible with code elements. This AST is then traversed to populate a semantic model of the code expressed in terms of concepts defined in an ontology we call Code Extraction Meta-model (CEM)[5]. The CEM defines concepts to represent various code structures, relationships among them, and other information useful in upstream processing. Some of the concepts included are: *CodeBlock, Class, Method, ConditionalBlock, LoopBlock, MethodCall, CodeVariable*. As expected in an ontology, some of these concepts are sub concepts of others. For example, *Method* is a subclass of *CodeBlock*. Each of these concepts have a set of properties to represent the relationships with other elements. For example, *Method* has the properties: *arguments*, *returnTypes*, *calls*, *isCalled*. It also inherits some properties from *CodeBlock*: *beginsAt*, *endsAt*, *serialization*, *containedIn*, among others.

[5] https://raw.githubusercontent.com/GEGlobalResearch/DARPA-ASKE-TA1/master/Ontology/M5/ExtractedModels/CodeExtractionModel.sadl

The CEM also includes a number of inference rules that determine whether variables referenced in a code block should be considered implicit inputs or outputs of the code block. This knowledge is crucial for enabling the composition of sub models into executable models. The semantic model of the code is enhanced by reasoning using these Jena inference rules.[6] For example, one rule looks at the references to a code variable within a method and identifies the first reference. Another rule looks at the usage of the code variable in the first reference and if the variable is an input to the statement or method call containing that reference then it must have been given a value outside of the method and so is an implicit input to the method.

Comments are given special attention. They are sent to *Text2Triples*, with the domain ontology as context, to extract semantic information that can be used to identify the semantic meaning of code variables. Currently, this extraction is focused on method inputs and outputs, but in the future, extraction can be done at a finer level of granularity to find interesting statements and code blocks and the semantic significance of their code variables. The enhanced code model is then queried to find methods of interest. Querying is accomplished using SPARQL (Harris, Seaborne, and Prud'hommeaux 2013) and interesting methods might be those that do computation, inferred by looking at the operators within the statements of a method.

Selected methods are serialized, with augmented semantic type information, and displayed to the user as extracted equations expressed in controlled English using the Semantic Application Design Language (SADL) tool (Crapo and Moitra 2013). Missing information is highlighted for user-assisted curation, either by editing the equation directly or by requesting additional extraction. Once an equation is complete – i.e., the meaning of inputs and outputs is described in domain terms – it is added to the KG along with the extracted method and its translation. Translation of the method to the target language is done before curation so that the user can, if desired, see the model in that language as well as the source language. Translation from Java to Python is accomplished via a slightly modified version of the open-source java2python[7] library. While *Code2Triples* is currently implemented for extraction from Java code, it is designed to be extensible with a view to support a variety of other programming languages in the future. Barring the code parser, *Code2Triples* is fairly language-agnostic.

### Text2Triples

The *Text2Triples* module extracts scientific concepts, equations, and augmented types from text. Additionally, it converts the extracted equations into executable Python code. We develop two parallel approaches to extract scientific concepts from text: a) Knowledge-driven Concept Extraction approach, and b) Data-driven concept and equation extraction approach. The motivation behind this two-pronged approach was to overcome any shortcomings of the data-driven model by leveraging user-specified domain concepts.

**Data-driven Concept and Equation Extraction**
The data-driven concept and equation extraction component leverages a supervised model to extract scientific concepts and equations from text. We treat the extraction problem as a named-entity recognition (NER) task and train a sequence tagging model focused on two entity types – concept and equation. We use the setup provided in the *flair* framework[8] to train the sequence tagging model. We stack pre-trained GloVe embeddings (Pennington, Socher, and Manning 2014) with task-trained character-level embeddings of words (Lample et al. 2016) as input to bidirectional LSTM (BiLSTM) with a conditional random field (CRF) decoding layer (Huang, Xu, and Yu 2015). The stacking process concatenates different embeddings to produce a single vector and has been shown to produce better performance in sequence tagging tasks (Akbik, Blythe, and Vollgraf 2018). The intuition behind the use of character embeddings is to help the model better parse equations in text which are composed of variable names consisting of one to few characters and mathematical operators. While the equations were manually annotated in our training and test data, the scientific concepts were automatically annotated via distant supervision. Details are outlined in the evaluation section.

**Knowledge-driven Concept Extraction**
The Knowledge-driven Concept Extraction component leverages a domain ontology provided by the user to identify scientific concepts in text. As a dictionary-based approach, it leverages existing scientific concepts already extracted by the system and corrected by users and/or concepts populated by them in a domain ontology. To extract concepts from text, we use UIMA ConceptMapper (Tanenblatt, Coden, and Sominsky 2010) along with the user provided domain ontology as the sources for the UIMA ConceptMapper dictionaries. This is accomplished by translating the ontology in RDF/XML format into UIMA ConceptMapper dictionaries in XML format. Each class, property in the domain ontology is translated into its canonical form and the variants for the dictionary format. The canonical form of a concept is represented by a uniform resource identifier (URI) of the concept, while its variants include the preferred name and synonyms of the concept.

**Merge and Align**
Leveraging multiple approaches to extract scientific concepts leads to duplication. Duplicate concepts are eliminated by simply performing a strict string similarity matching between the names of concepts extracted by the data-driven and knowledge-driven approaches. While this approach seems naive, it works well in the scientific domain which does not consist of too many ambiguous concept names.

---

[6] https://jena.apache.org/documentation/inference/index.html
[7] https://github.com/natural/java2python

[8] https://github.com/flairNLP/flair

Once the list of concepts is merged, each concept is aligned with concepts appearing in Wikidata (Vrandečićm and Krötzsch 2014). Aligning extracted concepts with external concepts in Wikidata allows our system to pull in additional information (e.g. units associated with a concept) which may not be readily available or which we may not have been extracted from the text that was processed. At the time of writing, Wikidata consisted of 89,067,076 items.[9] To simplify the alignment problem and support real time extraction and alignment of scientific concepts, we extracted a subset of Wikidata (concepts URIs and all their English language labels) and loaded it into a local Elasticsearch[10] index. Elasticsearch is a fast text search engine built on top of Lucene and provides REST APIs to execute search queries. We recursively include all concepts that are an instance of or sub class of the concept "physical property"[11] on Wikidata to narrow our focus to concepts appearing in the physical scientific domain.

Given that entity disambiguation is less of a challenge in the scientific domain, we generate a candidate set of Wikidata concepts for an extracted concept by leveraging Elasticsearch's match query. The match query performs fuzzy full-text search over the index, comparing the extracted concept text with the Wikidata concept labels. These candidates are further re-ranked by computing the Sørensen–Dice coefficient score between the extracted concept's text and the Wikidata label. The top-ranked candidate (if the score is >= 0.9) is selected as the Wikidata alignment.

**Generating Executable Code from Text Equations**

This component converts the text equation extracted by the Data-driven concept and equation extraction module into a piece of Python code which represents a function that can be run by a regular Python interpreter.

The component is essentially a rule-based parser scanning the input text string from left to right and creating the equivalent code as it does so. The parser works by processing the input text string one character at a time. Based on whether the character is alphabetic (a-z, A-z), numeric (0-9), operator (+, -, /, *, ^) or period (.), it determines what to do next. It also considers the next character and the previous character in making these determinations. For example, if there is a '2' followed by a '.', this could mean it is the beginning of a floating-point number '2'. If it is a '2' followed by a ')', this could mean it is the end of an expression and in that case, it has to backtrack to find the matching '(' that marks the start of the expression. For more than one bracket like character, e.g., "2 –b ) }", the parser has to backtrack until it finds all matching closing brackets. All bracket like characters are converted to ')' and '(' since those are the legitimate characters in Python. Sets of characters such as "exp" and "log" are not treated as simply characters but as mathematical operators and converted appropriately. Our parser currently recognizes a fixed set of mathematical operators.

While the input is a single string that represents an equation in text format, the module's output has several pieces of information. In addition to the converted Python code function, it also returns the extracted left-hand-side and right-hand-side expressions, the original input text string, and where applicable, multiple code interpretations of the text string. For instance, if the left hand side has a mathematical operator, e.g., "a * b = c + d", the code will return two equations – one corresponding to "a = (c + d) / b" and the other corresponding to "b = (c + d) / a". The output also includes a list of input variables and return/output variable for the equation/generated Python function.

**Extracting Augmented Types for Equation Variables**

Equations extracted from text (and the accompanying Python methods) are further enhanced by extracting the semantic meaning of its variables, i.e., their augmented types. The *Augmented Types Extraction* component extracts the semantic meaning for each variable in an equation by leveraging the context provided by the surrounding text. Consider the following example text:

> "An analysis based on conservation of mass and momentum shows that the speed of sound a is equal to the square root of the ratio of specific heats g times the gas constant R times the temperature T.
>
> a = sqrt [g * R * T]
>
> Notice that the temperature must be specified on an absolute scale (Kelvin or Rankine)"

In this case, the semantic meaning *speed of sound* of the equation variable *a* is captured in the line above the equation. In other cases, this evidence might be present a few lines above or below, but often close to where the equation appears. We use this knowledge to develop a rule-based approach to extract the augmented types for equation variables. Typically, limited patterns are used to define equation variables in text such as "mass m", "m is mass", or "mass is defined as m" to name a few. We combine semantic information (e.g. extracted concepts) along with textual patterns to develop our rules. These rules are applied to +/- 3 lines above and below every extracted equation. An example of such a rule includes searching for the pattern SCI_CONCEPT TOKEN (e.g. temperature T, where temperature is tagged as concept by data or knowledge-driven extraction) and assigning the concept (temperature) as the token's (T)

---

[9] https://www.wikidata.org/wiki/Wikidata:Statistics
[10] https://www.elastic.co
[11] https://www.wikidata.org/wiki/Q4373292

augmented type if the token is a input/output for the equation.

**Ontology for Knowledge Extracted from Text**

Scientific concepts, models, and augmented types extracted from text sources are also captured and organized in the knowledge graph using an ontology with a number of abstract concepts such as *ScientificConcept* and its subclass *UnittedQuantity*. Naturally, we use the latter extensively as most scientific properties are measurable and have units, which are represented using properties *value* and *unit* of this class. This ontology also includes an *Equation* class and its subclass *ExternalEquation*. These are used to represent atomic scientific models. The former represents models for which the system extracted and stored the implementation code, while the latter represents models whose implementation lies outside the system (e.g. accessed via a REST API). The main properties of these classes represent a model's *arguments* and *returnTypes*. Arguments (model inputs) are of type *DataDescriptor*, which associates to an argument a name, a *datatype* (e.g. float), and an *augmentedType*. The augmented type is the important property here, as it describes the meaning of the input variable in scientific domain terms. Typically, the *augmentedType* property will link an argument to a subclass of *UnittedQuantity* (e.g., *Temperature* or *Speed*) and a list of acceptable units. The *returnTypes* of a model associate similar information with the outputs of the model.

**Controlled-English Interface**

SADL and its grammar lets users create and edit OWL ontologies in a controlled-English vocabulary. We extend the open-source SADL tool to allow interactive dialog between human and the KG system. We believe such an interface will let domain scientists easily consume or update knowledge.

# Evaluation

We evaluate our system against a dataset of code and text sources from the NASA's Hypersonic Aerodynamics website[12]. The website has a collection of pages aimed towards undergraduates on several topics related to the hypersonics domain. Each page describes a scientific topic or phenomenon (e.g. Speed of Sound, Isentropic Flow) illustrated with graphics and mathematical equations. A subset of these pages are interactive in nature, allowing users to download and run simulations via Java applets. We use the source of these Java applets as the dataset for evaluating *Code2Triples*. We evaluate the performance of *Code2Triples* against 8 applets and the performance of *Text2Triples* against concepts and equations appearing in 63 pages.

**Code2Triples**

The applets in our dataset had few comments that were directly identifiable with code elements (no Javadoc comments were present) and intermixed scientific computation and graphical user interface (GUI) element. We eliminated GUI elements by specifying classes to ignore in the code meta-model. The expected number of relevant computational methods to be extracted from these 8 applets were 132 (ranging from 6 to 29 in any single applet). Our system was able to successfully extract all 132 methods. Because of the lack of comments, no augmented type information was extracted. However, for two applets, we processed text from the associated webpages to extract augmented types for code variables. We were able to extract correct augmented types for 3 out of 8 variables. Lack of consistency between the variable names in code and text, absence of code comments, and bad variable naming posed difficulties in achieving better performance.

**Text2Triples**

We use the notion of "distant supervision" (Mintz et al. 2009) to automatically annotate our corpus of webpages with BIO tags for the entity type "scientific concept" to train our data-driven model for concept and equation extraction. We create an OWL (McGuinness and Van Harmelen 2004) ontology of scientific concepts by extracting the titles for Wikipedia pages that are assigned the category of "Physical Quantities[13]". Our ontology consists of 223 concepts, each assigned a unique URI and one alias/synonym (the title of the Wikipedia page). Example concepts from the ontology include *Amplitude*, *Angular acceleration*, *Mass,* etc. We automatically translate this ontology into an XML format dictionary and use UIMA ConceptMapper (as described in the Knowledge-driven Concept Extraction subsection) to annotate our corpus. Concepts extracted by UIMA ConceptMapper are assigned the appropriate 'B' and 'I' tags while rest of the tokens are assigned the 'O' tag. The data-driven model is also trained to identify the entity type "equation". The 'B' and 'I' tags for equations in our corpus are added manually. The annotated dataset is randomly shuffled and split into 2/3rd parts for training and the remaining for test. The training dataset consisted of 913 concepts, 295 equations whereas the test dataset consisted of 484 concepts, 123 equations.

We use the flair framework to train a BiLSTM-CRF sequence tagging model (described in Data-driven Concept and Equation Extraction subsection). The model is trained with 256 hidden states, a learning rate of 0.1, batch size of 32 and the maximum epochs were kept to 150. The framework used micro F1 score as the evaluation method. Flair employs a simple learning rate annealing method in which the learning rate is halved if training loss does not fall for 5 consecutive epochs. It chooses the model that gives the best F-measure in the best epoch as judged over the test data. It took 95 epochs to train the model. The model performance over the test dataset is show in Table 1. The results are reported on a single run.

---

[12] https://www.grc.nasa.gov/www/BGH/shorth.html

[13] List of Wikipedia pages under the "Physical quantities" category: https://en.wikipedia.org/wiki/Category:Physical_quantities

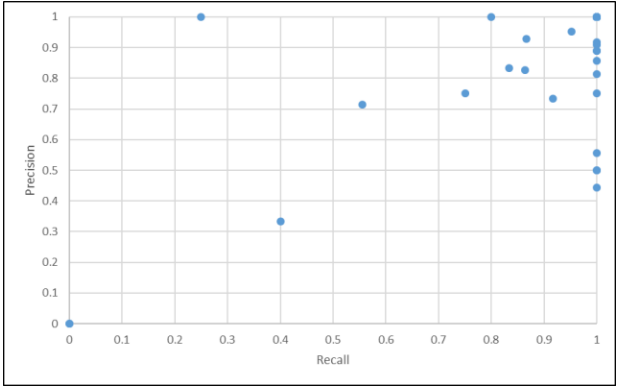

*Figure 2. Precision Recall values for each of the 63 pages. A dot represent a (recall value, precision value) pair for a given page*

The key information we wish to extract from these webpages are the equations appearing in the text and converting them into executable Python code to be stored in the knowledge graph. We present "page" level evaluations, since from an overall system point of view, we wish to augment the knowledge extracted from code with additional knowledge from text sources. In our current dataset, every Java applet is associated with one webpage providing additional context to understand the concepts captured in the code. Thus, it becomes important to accurately extract as much as knowledge the system can from a given page.

|  | Precision | Recall | F-score | Accuracy |
|---|---|---|---|---|
| Concept | 0. 9761 | 0. 9109 | 0. 9424 | 0. 9424 |
| Equation | 0. 9322 | 0. 8871 | 0. 9091 | 0. 9091 |

*Table 1. Model's performance over the test dataset*

We do a deeper evaluation of the *Text2Triples*' ability to accurately extract equations from 63 pages under the NASA Hypersonic Aerodynamics website. We manually create a ground truth dataset of 358 equations from these 63 pages and compute two metrics: **precision** – which measures what fraction of the extracted equations are relevant/correct and **recall** – which measures the fraction of expected equations that were retrieved/extracted by the system. The **average precision** over the 63 pages is **0.89** and the average recall is **0.92**. Figure 2 gives a breakdown of the precision, recall values for the individual pages (each page can consist of several equations). As can be seen from the graph, most pages have both a high precision and recall value simultaneously, indicating that the system was not only able to extract all expected equations but also most of the extracted equations were correct. In certain cases, the system also extracted additional incorrect/irrelevant equations which accounts for cases that have high recall values with lower precision values. Examples of incorrect, irrelevant and incomplete equations include "$v = .814$ cubic meters/kilogram", "$5^2 = 3^2 + 4^2$", and, "$x) = r * u * (del\ u\ /\ del\ x)$".

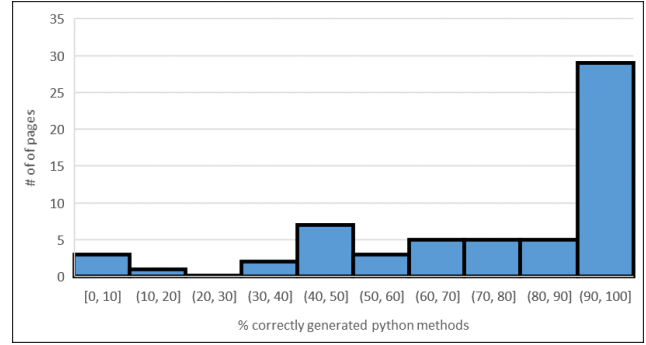

*Figure 3. Histogram showing the distribution of accuracy of generated python methods across different pages*

We also evaluate the system's ability to convert extracted textual equations into executable Python code. While there are **358 equations in our ground truth dataset**, as mentioned above, the **system extracts certain incorrect/irrelevant equations** leading to **374 equations** that the system attempted to translate into executable Python code. Out of 374 equations, our text to python module failed to generate any Python code for 98 of those. Some of those failed conversions can be attributed to the partial/incorrect extraction of equations from text. Given the nature of the equations that appear in textual format, the system often converts an equation into multiple Python methods. For example, for the equation "$E2 - E1 = Q - W$", Text to Python will generate two Python methods – one with E2 as the return variable (equation interpreted as "$E2 = Q - W + E1$") and one with E1 as the return variable (equation interpreted as "$E1 = -Q + W + E2$"). Overall, from the remaining 276 text equation strings, text to Python generated 374 executable Python methods across 60 pages, of which **256 were deemed correct (77.7%)** by our subject-matter-expert.

Figure 3 gives a distribution of the accuracy of generated Python methods across different pages. As seen in the figure, accuracy for 39 out of 60 pages is 70% or greater with almost **50% of the pages having an accuracy of 90%** or greater. Simple equations are the easiest to convert into functional Python code. Examples of such equations would be where there is a single variable on the left hand side and simple operations such as +, - , *, / on the right hand side. Slightly more complex equations, such as those having an operator on the left-hand side, were also handled without too many issues. The complicated ones were usually the ones that involved the exponent operations, e.g., ^ or e^. The tricky part was to backtrack from the ^ to figure out to what text segment the ^ operator was being applied. Especially complicated cases include having to deal with the unary negation operator in addition, e.g., e^-(1 + gamma) or e^-T. Currently, the translation from text to Python assumes everything is literal text. Therefore, expressions such as "exp" or "degrees" occurring in the text string are being translated as is, i.e., "exp" and "degrees". The code generator is essentially a parser that works with whatever it knows about. Therefore, whenever it encounters a new condition that it has not seen before, it fails.

## Related Work

We divide this section by summarizing work related to different components in our system. Constructing semantically grounded knowledge representations from code has been done before (Atzeni and Atzori 2017), not necessarily for scientific model extraction. Graph4Code (Abdelaziz et. al. 2020), a knowledge graph of > 2 billion triples constructed from GitHub-based code and Stack Overflow posts seeks to improve coding practices via auto-extraction of Python classes and methods. Our system enables construction of KGs from code for different scientific domains, and also emphasizes extraction from equations and comments in text to further augment the KGs. For scientific code, existing approaches to construct semantic KGs include both supervised (Jiomekong et. al. 2019) and unsupervised learning-based (Cao and Fairbanks 2019) techniques. Our rule-based reasoning approach does not require any training and can also consider implicit inputs and outputs in code. Hein (2019) presents a program analysis approach that translates source code into an intermediate structured dataflow form prior to scientific model extraction.

Named entity recognition (NER) is a well-researched area. State-of-the-art systems are evaluated on typical entity types such as person, place, and organization (Yadav and Bethard 2018). NER systems have also been developed for specific scientific domains such as medical (Xu et al. 2017), biomedical (Lee et al. 2019) and computer science (Beltagy, Lo, and Cohan 2019) to name a few. To the best of our knowledge, no specific NER model is available to extract concepts related to physics-based systems. A NER system developed for the astronomy literature (Murphy, McIntosh, and Curran 2006) would come close, but the model is trained on extracting specific sub entity types (e.g. galaxy, ion, telescope) within the domain.

Limited work has focused on extracting equations or mathematical expressions that appear in text or scientific publications. Most research in this space has been dedicated towards leveraging optical character recognition along with other image processing techniques that use image-based features such as fonts and styles to extract equations (Phong et al. 2020), (Zhang, Du, and Dai 2018), (Iwatsuki et al. 2017), (Garain and Chaudhuri 2007), (Chan and Yeung 2000).

Equation extraction from textual sources focuses more on processing structured data such as LaTeX (Chen et al. 2014) and MathML (Cui et al. 2011) appearing on webpages. Limited work has also focused on extracting mathematical expressions from textual descriptions (Kristianto and Aizawa 2014) or from simple and often partial formulae embedded in a mathematical problem solving question (e.g. *Find arithmetic mean of the numbers in the list $8 - a, 8, 8 + a$*) (Fernando, Ranathunga, and Dias 2019).

Like the proposed approach in this paper, Tian et al. (2017) focus on extracting equations from unstructured text. They develop a Hidden Markov model capturing eight states to represent an equation. Additionally, they leverage a dictionary of trigonometric functions as a set of trigger words in the model. Compared to Tian et al., our proposed approach requires no hand-crafted feature engineering or pre–curated dictionary of mathematical functions. The code and dataset used in their paper was not publicly available to evaluate the complexity of the equations handled by their approach or to evaluate our approach on their dataset to compare the two.

## Conclusion and Future Work

Scientific models, when suitably annotated with semantically grounded information about their inputs and outputs, can enhance their discoverability and understandability by domain experts within various scientific domains. Further, a machine-interpretable knowledge graph (KG) linking such computable scientific models (CMs) will facilitate their automated composition to assist with complex diagnostic and prognostic queries. We presented a system for the automated creation and human-assisted curation of such a KG of CMs and applied it to the hypersonic flight scientific domain.

Our system employs a multi-pronged, modular approach to extract scientific models from both source code as well as text-based sources, and to populate executable instances of these models in a KG. The *Code2Triples* module employs a semantic reasoning approach to identify and represent relevant code artifacts and align them against a meta-model of scientific code. In conjunction, *Text2Triples* augments this information via a combined knowledge- and data-driven approach to extract semantically grounded concepts and equations from inline comments and external documentation. Our system's controlled-English interface allows domain experts to both interactively query as well as curate the KGs.

While experiments on datasets from NASA's Hypersonics website have shown promise, we identify some potential areas of improvement for our system in the future: *Code2Triples* currently works at the method level and only on Java code. Two extensions are desirable: 1) to reason and perform model extraction at finer levels of granularity, such as identifying statements or code blocks as being of significance, and 2) to demonstrate that our system can be extended to other programming languages by implementing the relevant AST parsers. Augmented types extraction in *Text2Triples* can be migrated from rules-based to data-driven approach by extending the existing NER to identify a variable and its semantic meaning along with units and default values wherever present.

# Acknowledgements

This work was supported by the Defense Advanced Research Projects Agency (DARPA) as part of the Automated Scientific Knowledge Extraction (ASKE) program under agreements no.: HR00111990006 and HR00111990007.